\documentclass[aps,twocolumn,showpacs]{revtex4}
\usepackage{dcolumn}
\usepackage{graphicx}
\usepackage{amsmath}
\usepackage{verbatim}
\usepackage{amssymb}
\usepackage{psfrag}
\usepackage{bm}
\usepackage{wrapfig}
\usepackage{subfigure}
\usepackage{makeidx}
\usepackage{bm}
\usepackage{setspace}
\usepackage{epsf}
\usepackage{multirow}
\usepackage[colorlinks,linkcolor=blue,urlcolor=blue,citecolor=blue]{hyperref}

\DeclareMathOperator{\sech}{sech}
\begin{document}
\title{Exact analytical soliton solutions of $N$-component coupled nonlinear Schr\"{o}dinger equations with arbitrary nonlinear parameters}
\author{Ning Mao}
\author{Li-Chen Zhao}\email{zhaolichen3@nwu.edu.cn}
\address{School of Physics, Northwest University, Xi'an, 710127, China}
\address{Peng Huanwu Center for Fundamental Theory, Xi'an 710127, China}
\address{Shaanxi Key Laboratory for Theoretical Physics Frontiers, Xi'an, 710127, China}

\date{\today}
\begin{abstract}
Exact analytical soliton solutions play an important role in soliton fields. Soliton solutions were obtained with some special constraints on the nonlinear parameters in nonlinear coupled systems, but they usually do not holds in real physical systems. We successfully release all usual constrain conditions on nonlinear parameters for exact analytical vector soliton solutions in $N$-component coupled nonlinear Schr\"{o}dinger equations. The exact soliton solutions and their existence condition are given explicitly. Applications of these results are discussed in several present experimental parameters regimes. The results would motivate experiments to observe more novel vector solitons in nonlinear optical fibers, Bose-Einstein condensates, and other nonlinear coupled systems.
\end{abstract}
\maketitle
\textit{Introduction}---Analytical soliton solutions can describe the dynamics of localized waves in many nonlinear systems well, and some of them have even been used to direct experimental observations \cite{1dark,2bright,3DB,4DBB,5Magnetic,6Magnetic,7rogue}. Analytical solutions usually contain exact solutions and approximate solutions.  Exact analytical soliton solutions have been paid much more attention due to their beauty and convenience for uncovering underlying physics \cite{8BS,9Peregrine,10K-M,11AhB,12CQds,13DB}. Unfortunately, they are usually given with some special constrain conditions  \cite{14Manakov,15Darb,16Darb,17inverse,18Hirota}. For example, most soliton solutions of multi-component coupled nonlinear systems were obtained with some special constraints on the nonlinear parameters, e.g., the Manakov model \cite{14Manakov}.   However, those special constraint conditions usually do not holds in real nonlinear coupled systems \cite{3DB,4DBB,5Magnetic,6Magnetic,19fiber,20SP}. It highly demands us to release the constrain conditions on nonlinear parameters to derive exact soliton solutions.

In this letter, we release all usual constrain conditions on nonlinear parameters for exact analytical vector soliton solutions in $N$-component coupled nonlinear Schr\"{o}dinger equations, which have played important roles in soliton fields due to their simplicity and wide applications in nonlinear systems \cite{19fiber,20SP,21Ncoupled}. Our attempt could motivate more efforts to derive exact soliton solutions of other nonlinear models \cite{22threewave,23soc,24couple-quntic}. Based on our proper trial functions and the modified Lagrangian variational method, we give the exact soliton solutions and their existence condition explicitly, which covers all previously known exact soliton solutions and predicts some new physical properties for vector solitons. As an example, we show two- and three-component vector soliton solutions and their existence region in detail. Applications of these results are discussed in  Bose-Einstein condensates with present experimental parameters regimes.

\textit{The exact vector soliton solutions and Lagrangian method}---Coupled nonlinear Schr\"{o}dinger equations have been used to describe nonlinear wave dynamics in many different physical systems \cite{21Ncoupled,25xde,26xde}, such as Bose--Einstein condensates \cite{26xde} and nonlinear optical fibers \cite{19fiber,27xde}. We consider a general $N$-component coupled nonlinear Schr\"{o}dinger equations with arbitrary nonlinear parameters, which can be written as
\begin{equation}\label{xde}
	\begin{split}
		{\rm{i}}\partial_t \bm{\Psi}=\big(-\frac{1}{2}\partial_{xx}+\sum_{j=1}^{N} g_{ij}|\psi_j|^2\big)\bm{\Psi},
	\end{split}
\end{equation}
where $\bm{\Psi}=(\psi_1,~\cdots,~ \psi_i,~\cdots,~\psi_N)^{T}$ and where $\psi_{j}$ stands for the wave function of the $j$th component. The nonlinear coefficient $g_{ij}$ is the intra-interaction in one component (inter-interaction between two components) for $i=j\ (i\neq j)$, and $g_{ij}=g_{ji}$. They admit scalar solitons ($N=1$) and vector solitons ($N>1$). The exact scalar bright and dark soliton solutions were  derived by the B\"{a}cklund transformation \cite{15Darb,16Darb}, inverse scattering method \cite{17inverse}, and Hirota bilinear method \cite{18Hirota} (noting that $N=1$ cases are always integrable). But there are many nonlinear parameters when $N>1$, which makes the coupled systems usually no longer integrable. When all these nonlinear parameters are equal, the model becomes the well-known integrable Manakov model \cite{14Manakov}, whose vector soliton solutions have been derived by many different methods \cite{15Darb,16Darb,17inverse,18Hirota}. The soliton solutions of the Manakov model have motivated many experiments to observe dark-dark (DD) solitons \cite{28DD E1,29DD E2}, dark-bright (DB) solitons \cite{3DB}, etc.

However, the constrain conditions on nonlinear parameters  usually do not holds in real nonlinear coupled systems \cite{3DB,4DBB,20SP,30F2,31DF2,32F2,33DF2}.  Many scientists tried to address the cases for which the special constraint conditions are violated, and some approximate analytic solutions were obtained \cite{34VD BB,35VD Guass,36VD Guass,37sech,38carr,39carr}. More approximate analytic solutions  could be given by the perturbation method \cite{40pertubation}, multiscale expansion method \cite{41multiscale,42multiscale}, and Lagrangian variational method \cite{43B,44B,45B,46VD,47VD,48VD}. We wish to find exact analytic solutions for more general conditions. Notably, some exact soliton solutions can still be derived by the Lagrangian variational methods \cite{49B,50VD}. This hints that it is possible to derive exact soliton solutions for nonintegrable cases and even more general cases, by developing the methods.

For the Lagrangian variational methods, the trial functions are critical for precision. The Gaussian profile is usually used for localized wave packets due to its simplicity and easy calculation. However, this form usually fails to obtain exact solutions in nonlinear systems. It was further suggested that the sech- (tanh-)type ansatz could be more accurate than the Gaussian ansatz \cite{34VD BB}. For bright or dark solitons in the $i$th component, we introduce the trial wave function as
\begin{eqnarray}\label{shitan}
		\psi_{iD}=\!\!\!\!&&\big(\textrm{i}\sqrt{a_i^2-f_i^2(t)}+f_i(t)\tanh\{w_i(t)[x-b(t)]\}\big)e^{\textrm{i}\theta_i(t)},\nonumber\\\\
		\psi_{iB}=\!\!\!\!&&f_i(t)\sech\{w_i(t)[x-b(t)]\}\textrm{e}^{\textrm{i}\{\xi_i(t)+[x-b(t)]\phi_i(t)\}},\nonumber
\end{eqnarray}
where $\psi_{iD}$ ($\psi_{iB}$) denotes the wave function of the dark (bright) soliton, $f_i$ and $w_i$ describe the amplitude and width of the dark (bright) soliton, respectively, $a_i$ is the background of the dark component, the central position of the soliton is $b(t)$, $\theta_i$ and $\xi_i$ are the time-dependent phases of the dark and bright components, respectively, and $\phi_i$ is related to the velocity of the bright soliton.

We introduce the Lagrangian density as $\mathcal{L}=\sum_{i=1}^N\big[\frac{\textrm{i}}{2}(\tilde{\psi_i}^*\partial_t\tilde{\psi_i}-\tilde{\psi_i}\partial_t\tilde{\psi_i}^*)
(1-\frac{a_i^2}{|\tilde{\psi_i}|^2}\delta_{D,i})-\frac{1}{2}|\partial_x\tilde{\psi_i}|^2 -\sum_{j=1}^N\frac{g_{ij}}{2}(|\tilde{\psi_j}|^2-a_j^2 \delta_{D,j})(|\tilde{\psi_i}|^2-a_i^2 \delta_{D,i})\big]$, where
\begin{equation*}
	\delta_{D,i}=\left\{
	\begin{array}{cl}
		0, & \tilde{\psi}_{i}\ \textrm{denotes\ bright soliton}, \\\\
		1, & \tilde{\psi}_{i}\ \textrm{denotes\ dark soliton}, \\
	\end{array}
	\right.
\end{equation*}
to obtain Euler-Lagrangian equations. In particular, the terms $1-a_i^2/|\tilde{\psi_i}|^2\delta_{D,i}$ and $|\tilde{\psi_i}|^2-a_i^2 \delta_{D,i}$ were introduced in the Lagrangian density for dark solitons \cite{50VD}. Notably, the modified $\mathcal{L}$ corresponds to dynamical equation ${\rm{i}}\partial_t \tilde{\psi}_i=-\frac{1}{2}\partial_{xx}\tilde{\psi}_i+\big[\sum_{j=1}^N g_{ij}(|\tilde{\psi}_j|^2-a_j^2 \delta_{D,j})\big]\tilde{\psi}_i$ $(i=1,\cdot\cdot\cdot,N)$. This is different from Eq. \eqref{xde}, but the solutions of the two dynamical equations can be transformed to each other by the phase factor $\psi_i/\tilde{\psi}_i=\textrm{e}^{\textrm{i}\theta_i(t)}$, where $\theta_i(t)=-\sum_{j=1}^Ng_{ij}a_j^2t \delta_{D,j}$. We obtain the Lagrangian ($L$) by substituting Eq. \eqref{shitan} into $\mathcal{L}$ and integrating over space from $-\infty$ to $+\infty$, which can be simplified as
\begin{widetext}
\begin{equation}\label{int}
		\begin{split}			L=&\sum_i^N\bigg\{\Big[2f_i^2(t)\phi_i(t)\frac{b'(t)}{w_i(t)}-2f_i^2(t)\frac{\xi_i'(t)}{w_i(t)}-\frac{1}{3}f_i^2(t)w_i(t)-f_i^2(t)\frac{\phi_i^2(t)}{w_i(t)}-\frac{2}{3}g_{ii}\frac{f_i^4(t)}{w_i(t)}\Big]
\tilde{\delta}_{D,i}+\Big\{-2f_i(t)\sqrt{a_i^2-f_i^2(t)}b'(t)\\&+2 a_i^2\arcsin\Big[\frac{f_i(t)}{a_i}\Big]b'(t)-\frac{2}{3}f_i^2(t)w_i(t)-\frac{2}{3}g_{ii}\frac{f_i^4(t)}{w_i(t)}\Big\}\delta_{D,i}-\sum_{j(j\neq i)}^Ng_{ij}f_i^2(t)f_j^2(t)G_{ij}\bigg\},~~~~~~ (\tilde{\delta}_{D,i}=|\delta_{D,i}-1|).
		\end{split}
    \end{equation}
\end{widetext}

Specifically, the soliton width parameter $w_i$ is set to be different for different components, which is in contrast to the previous trial functions for vector solitons \cite{34VD BB,35VD Guass,36VD Guass,37sech,38carr,39carr}. In most previous studies, the trial functions were chosen as the Gaussian (sech- or tanh-type) ansatz for the unequal (equal) width parameter setting \cite{34VD BB,35VD Guass,36VD Guass,37sech,38carr,39carr}, partly because the Gaussian ansatz with different widths is much easier to calculate than the sech (tanh) type. This makes the Lagrangian variational results only give an approximate solution for the vector model \cite{34VD BB,35VD Guass,36VD Guass,37sech,38carr,39carr}. However, one encounters the problem that it is difficult to accommodate the factor $G_{ij}=\int^{+\infty}_{-\infty}\big(\pm \sech^2\{w_i(t)[x-b(t)]\}\big)\big(\pm \sech^2\{w_j(t)[x-b(t)]\}\big)dx$, and the sign $-~(+)$ corresponds to the dark (bright) component. We take these width parameters independently when deriving the Euler-Lagrangian equations. After obtaining the equations of motion by the Euler-Lagrangian formula, we finally calculate $G_{ij}$, $\frac{\partial G_{ij}}{\partial w_i}$, and $\frac{\partial G_{ij}}{\partial w_j}$ by setting $w_i=w_j=w$ (see details in Supplemental Material \cite{51M}). These operations bring us more constraint conditions on soliton parameters and finally enable us to obtain exact analytical soliton solutions.

Based on all simplified Euler-Lagrangian equations, we have the essential constraint equations of $f_i$ and $w$ as
\begin{eqnarray}\label{solution}
	\left(
	\begin{array}{ccccc}
		g_{11}&\cdots g_{1j}\cdots&g_{1N}&1 \\
		\vdots&\vdots&\vdots&\vdots&\\
		g_{i1}&\cdots g_{ij}\cdots&g_{iN}&1 \\
		\vdots&\vdots&\vdots&\vdots&\\
		g_{N1}&\cdots g_{Nj}\cdots&g_{NN}&1 \\
	\end{array}
	\right)\left(
	\begin{array}{cc}
		\pm f_1^2\\ \vdots\\ \pm f_i^2 \\ \vdots\\ \pm f_N^2\\w^2
	\end{array}
	\right)=0.
\end{eqnarray}
The sign $-~(+)$ corresponds to the dark (bright) component. The other parameters can be obtained as $b(t)=vt$, $\phi_i=v$, $\xi_i(t)=\frac{1}{2}(w^2+v^2)t+\theta_i(t)$, and the soliton velocity $v=w\sqrt{a_i^2-f_i^2}/f_i$.
The constraint equation on backgrounds of dark components is $a_i/a_j=f_i/f_j$ (for details, see \cite{51M}). In this way, we obtain many exact analytic vector soliton solutions when the nonlinear parameters can be arbitrary. Different types of vector solitons usually exist in different regions in nonlinear coefficient space. Their existence regions can also be clarified by the constraint conditions on soliton parameters. The above solution can cover all previously reported vector solitons for integrable cases ($g_{ij}$s are all equal) \cite{13DB,52BB2,53BBDD,54DDDB,55qin,56DDBpla,57DDD,58DBBN,59BBN,60BDN,61DDN,62DDN}.
We choose two- and three-component systems as examples to show the variational result explicitly based on vector soliton experiments in Ref. \cite{3DB,4DBB,5Magnetic,6Magnetic}.

\begin{figure}[http]
	\centering
	\includegraphics[width=86mm]{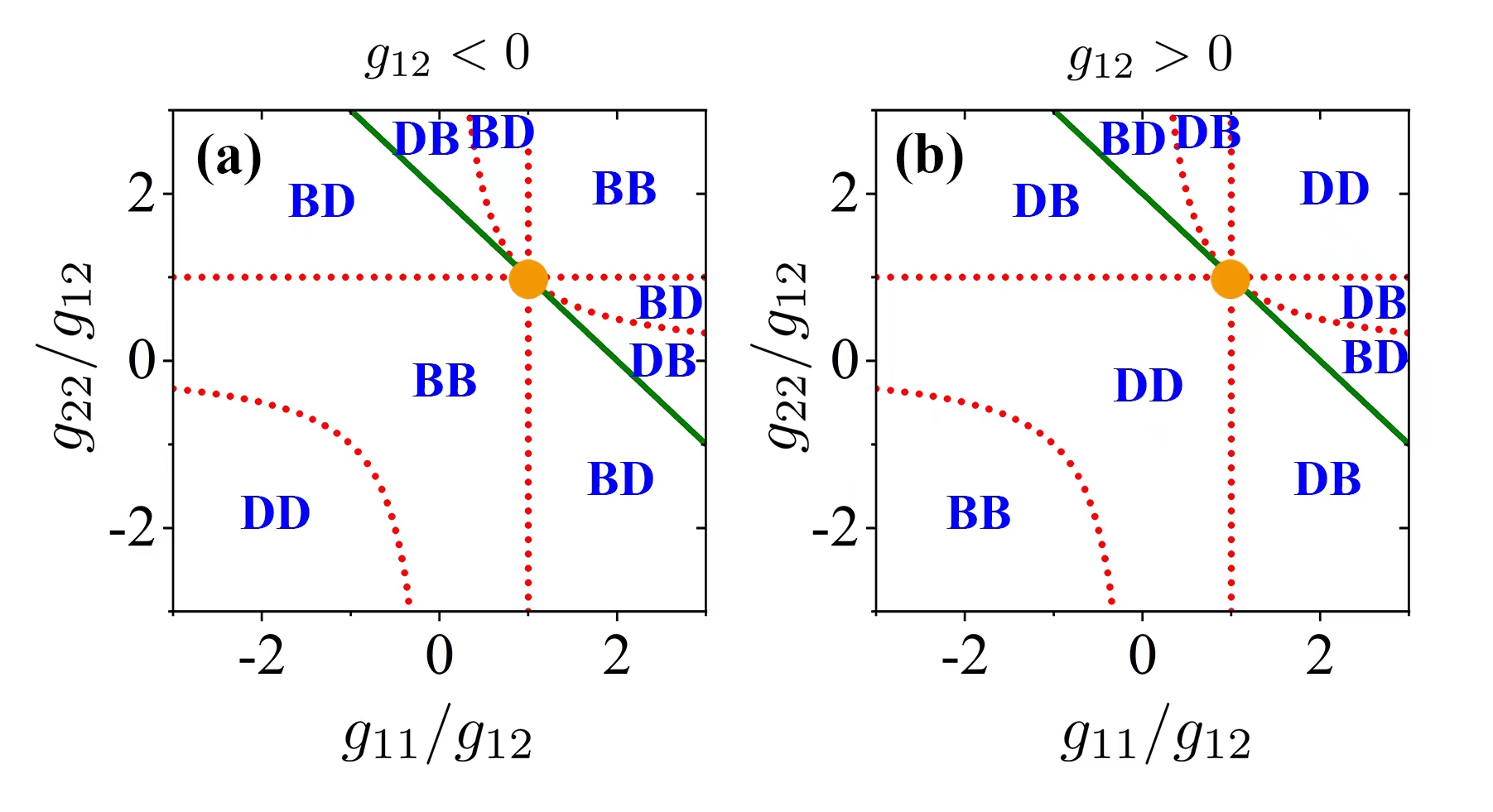}
	\caption{Phase diagram for different vector solitons in nonlinear coefficient space. (a) and (b) represent the cases of $g_{12}<0$ and $g_{12}>0$, respectively. The red dashed lines are the boundary lines for different vector soliton regions. The green lines denote the condition for spin solitons. The orange dot denotes the usual integrable case. Our variational method greatly extends the existence region for exact vector soliton solutions.}\label{zong}
\end{figure}

\textit{Two-component coupled systems}---For two-component systems ($N=2$), the condition for vector soliton solutions can be derived from Eq. \eqref{solution}, which is simplified as
\begin{equation}\label{region}
	\alpha_2 f_2^2=\frac{g_{11}-g_{12}}{g_{22}-g_{12}}\alpha_1 f_1^2,~w^2=\frac{g_{12}^2-g_{11}g_{22}}{g_{22}-g_{12}}\alpha_1 f_1^2,
\end{equation}
where $\alpha_i=\pm1$ and where the sign $-$ ($+$) is chosen for the dark (bright) soliton. Then, we clarify the existence regimes for different vector solitons by analyzing the soliton parameters of Eq. \eqref{region}. The phase diagram for different vector solitons is summarized in Fig. \ref{zong} ($g_{12}\neq0$), in which (a) and (b) show the results for the $g_{12}<0$ and $g_{12}>0$ cases, respectively. When $g_{12}=0$, the model decouples into two scalar models. Our variational method extends the existence region of exact vector soliton solutions from the ``orange dot" (Manakov model) to the whole parameter plane. The vector solitons can still be classified into four families, similar to the integrable model \cite{13DB,52BB2,53BBDD,54DDDB,55qin}. However, there are many additional constraints on the soliton parameters and nonlinear parameters, which are absent for integrable models \cite{13DB,52BB2,53BBDD,54DDDB,55qin}.

For bright-bright (BB) solitons, Eq. \eqref{region} gives the conditions $\frac{g_{11}-g_{12}}{g_{22}-g_{12}}>0$ and $\frac{g_{12}^2-g_{11}g_{22}}{g_{22}-g_{12}} >0$ on nonlinear parameters. Then, the region for the BB soliton can be given in Fig \ref{zong}. The region for DD solitons can be given in a similar way. We note that there is an additional constraint on the background amplitudes for the DD soliton ($a_2/a_1=f_2/f_1$) in nonintegrable cases. This is in sharp contrast to the Manakov model, for which there is no constraint on the ratio of background amplitudes for DD solitons \cite{53BBDD,54DDDB}.

For dark-bright (DB) or bright-dark (BD) solitons, the regions for them are shown in Fig. \ref{zong}. The DB and BD solitons are defined by the total density of the two components, which admits a dip (for DB) or hump (for BD). In particular, the total density can be uniform when $2g_{12}=g_{11}+g_{22}$, and the vector soliton in this case becomes the spin soliton \cite{63spin} (denoted by green lines in Fig. \ref{zong}). Note that the amplitude of the bright soliton can be larger than the dark soliton's background amplitude; i.e., BD soliton can exist even though the nonlinear interactions are all repulsive (all nonlinear parameters are positive). This characteristic is absent for the Manakov case \cite{13DB}, for which only DB soliton exists for repulsive interactions. Similarly, DB soliton can still exist with the nonlinear interactions all attractive (all nonlinear parameters are negative) for nonintegrable cases. Our DB soliton solution with zero velocity can be reduced to the ones given in Ref. \cite{64Vary1}.
The exact BB and BD (DD and DB) soliton solutions cannot coexist in the given nonlinear parameters for the nonintegrable cases (see Fig. \ref{zong}), in contrast to the integrable case \cite{13DB,52BB2,53BBDD,54DDDB,55qin}.

Our variational results can be reduced to a previously known exact soliton solution in the two-component model \cite{13DB,52BB2,53BBDD,54DDDB,55qin,63spin}. The general exact vector soliton solutions of two-component cases are provided explicitly in \cite{51M}, which partly overlap with the ones given by the periodic wave expansion method \cite{65puhan,66Jo}. We emphasize that our variational method here can be extended more conveniently to arbitrary $N$-component cases. Experiments have been performed on three-component systems and even five-component Bose--Einstein condensates \cite{4DBB,20SP,30F2,31DF2,32F2,33DF2}, but the exact analytic soliton solutions for them are still absent. These results motivate us to derive soliton solutions explicitly for more components cases. Next, we apply our variational result to three-component systems.

\textit{Three-component coupled systems}---The exact soliton solution of three-component systems has been widely studied for integrable models \cite{56DDBpla,57DDD,58DBBN,59BBN,60BDN}. Our variational result greatly widens the existence region of vector soliton solutions, which is very meaningful for soliton experiments \cite{4DBB}. The essential conditions for soliton solutions can be given as
\begin{equation}\label{three}
	\alpha_2 f_2^2=\frac{P_{123}}{P_{213}}\alpha_1f_1^2,~\alpha_3 f_3^2=\frac{P_{132}}{P_{213}}\alpha_1f_1^2,~w^2=\frac{Q}{P_{213}}\alpha_1f_1^2,
\end{equation}
where $Q=2g_{12}g_{13}g_{23}-g_{12}^2g_{33}-g_{13}^2g_{22}-g_{11}(g_{23}^2-g_{22}g_{33})$, $P_{lmn}=g_{ln}^2-g_{ll}g_{nn}+g_{lm}(g_{nn}-g_{ln})+g_{mn}(g_{ll}-g_{ln})$, ($l,m,n=1,2,3$), and $\alpha_i=\pm 1$ ($i=1,2,3$).
The exact soliton solutions can be given from Eq. \eqref{solution} with $N=3$, whose explicit expressions are given in \cite{51M}. The vector soliton can be classified into four families, and their existence conditions are summarized in Table \ref{table}.

\begin{table}[tp]
	\begin{spacing}{1.35}
	\begin{tabular}{c|ccc}
		\hline\hline
		Soliton type & \multicolumn{3}{c}{Parameter region} \\ \hline
		& ~$P_{123}/P_{213}$~&~ $P_{132}/P_{213}$~    &~  $Q/P_{213}$~    \\ \hline
		BBB&     $+$&     $+$&   $+$  \\
		DBB&     $-$&     $-$&   $-$   \\
		DDB&     $+$&     $-$&   $-$    \\
		DDD&     $+$&     $+$&   $-$      \\ \hline\hline
	\end{tabular}
    \end{spacing}
	\caption{Parameter conditions of different soliton solutions for the three-component coupled systems; ``$+$" and ``$-$" denote the parameters being positive and negative, respectively. }\label{table}
\end{table}

We cannot give explicit conditions of existence for different soliton solutions, as done in the two-component cases, since there are many more nonlinear parameters in the three-component cases. To show the physical meaning of the conditions in Table \ref{table}, we discuss the existence conditions of different vector solitons with setting $g_{12}=1,~g_{13}=2,~g_{23}=3$. With the repulsive interspecies interaction, the exact bright-bright-bright (BBB) soliton solution can still exist when $g_{11}<0,~g_{22}<1/g_{11} $ and $g_{33}<(-12+9g_{11}+4g_{22})/(-1+g_{11}g_{22})$. By further tuning the intra-interactions, we can obtain the existence region for dark-bright-bright (DBB), dark-dark-bright (DDB) and dark-dark-dark (DDD) solitons.

\textit{Applications in experiments}---We first discuss the applications of our solutions in a two-component Bose-Einstein condensate of $^{87}$Rb \cite{3DB,67DBpla} with hyperfine states $|1,-1\rangle$ and $|2,0\rangle$ (denoted by $\psi_1 $ and $\psi_2 $). The scatting lengths are $a_{11}=100.86a_0,\ a_{12}=98.98a_0,\ a_{22}=94.57a_0$, where $a_0=5.29\times 10^{-11}$ m is the Bohr radius.
The nonlinear parameters in our rescaling model satisfy $g_{11}:g_{12}:g_{22}=1:0.981:0.938$, and $g_{11}=0.0114$. The dynamical equations in the mean-field approximation become nonintegrable. The theoretical analysis of the experimental results is usually performed based on an integrable model, ignoring the small differences between the nonlinear parameters \cite{3DB}. This is partly because exact analytical solutions are very rare with the parameters in real experiments. We discuss the properties of vector solitons based on our variational results. With the real nonlinear parameters and identical atom density $5.8 \times 10^{13} \textrm{cm}^{-3}$ \cite{3DB}, our solution predicts that dark solitons exist only in the $\psi_1$ component and that they cannot exchange with the second component, in contrast to the prediction of the integrable model. The sound speed is predicted to be $0.766$ mm/s by our soliton solutions, which is smaller than the sound velocity $1$ mm/s given by the integrable results \cite{3DB}. Moreover, the constraint conditions on nonlinear parameters predict that there is no exact analytical DD soliton solution in this case but that the DD soliton solution can be given approximately by the integrable model. These characteristics may inspire experiments to check these differences.

There are many theoretical works in three-component systems \cite{56DDBpla,57DDD,58DBBN,59BBN,60BDN}. DDB and DBB solitons were realized experimentally in \cite{4DBB}. They obtained the approximate solution of DDB and DBB solitons by a multiscale expansion method. With the interaction parameters $g_{11}=g_{12}=g_{23}=g_{33}=g_1,~g_{22}=g_2,~g_{13}=g_3$ and $0<g_1<g_2<g_3$ \cite{4DBB}, we have $P_{123}/P_{213}>0,~P_{132}/P_{213}>0,~Q/P_{213}<0$. We predict that only an exact DDD soliton solution exists from Table \ref{table} if we ignore the particle transition effects in the experiment. The constraint on amplitude and width is derived as $f_2^2=(g_3-g_1)/(g_2-g_1)f_1^2$, $f_3=f_1$, and $w^2=(g_2g_3+g_1g_2-2g_1^2)/(g_2-g_1)f_1^2$.
These results predict that the first and third components always admit identical dark solitons due to the symmetry of the nonlinear parameters. If we tune the nonlinear parameters to satisfy $g_1\leq 0,~g_2<g_1$ and $g_3<g_1$, the exact BBB soliton solution can be obtained. Many other vector solitons can also be obtained in similar ways.

\textit{Conclusion}---Our work successfully releases all usual constrain conditions on nonlinear parameters for exact analytical vector soliton solutions in $N$-component coupled nonlinear Schr\"{o}dinger equations, which have played important roles in soliton fields. Our explicit soliton solutions and their existence condition could motivate  experiments to observe vector solitons in Bose-Einstein condensates \cite{4DBB,5Magnetic,6Magnetic,20SP} and nonlinear optical fibers \cite{19fiber,27xde}, especially when the nonlinear parameters deviate from the usual integrable conditions. They could provide an important supplement for the experimental observation of vector solitons, which are usually around integrable conditions \cite{3DB,4DBB,5Magnetic,6Magnetic}. The general vector soliton solutions are helpful to discuss the soliton dispersion relation and soliton transport for more general cases \cite{63spin}. Our attempt will stimulate more efforts to derive exact soliton solutions of other nonlinear models, which can describe real nonlinear systems better.

\section*{Acknowledgments}
This work is supported by the National Natural Science Foundation of China (contract nos. 12022513, 12047502) and the Major Basic Research Program of Natural Science of Shaanxi Province (grant no. 2018KJXX-094).

\end{document}


\title{The supplemental material of ``Exact analytical soliton solutions of $N$-component coupled nonlinear Schr\"{o}dinger equations with arbitrary nonlinear parameters"}
\author{N. Mao}
\author{L.-C. Zhao}
\address{$^{1}$School of Physics, Northwest University, Xi'an 710127, China}
\address{$^{2}$Peng Huanwu Center for Fundamental Theory, Xi'an 710127, China}
\address{$^{3}$Shaanxi Key Laboratory for Theoretical Physics Frontiers, Xi'an 710127, China}


 \begin{abstract}
In this supplemental material, we provide the details for deriving the exact soliton solutions of $N$-component coupled nonlinear Schr\"{o}dinger equations with arbitrary nonlinear parameters in Sec. \ref{1}. The explicit expressions of exact vector soliton solutions for two- and three-component coupled model are provided in Sec. \ref{2} and Sec. \ref{3}, respectively.
\end{abstract}
\date{\today}

\maketitle

\setcounter{figure}{0} \renewcommand{\thefigure}{S\arabic{figure}} %
\setcounter{equation}{0} \renewcommand{\theequation}{S\arabic{equation}}

\section{The derivation of N-component vector soliton solutions}\label{1}

We consider a general $N$-component coupled nonlinear Schr\"{o}dinger equations, which can be written as
\begin{equation}\label{xde1}
	\begin{split}
		{\rm{i}}\partial_t \bm{\Psi}=\big(-\frac{1}{2}\partial_{xx}+\sum_{j=1}^{N} g_{ij}|\psi_j|^2\big)\bm{\Psi},
	\end{split}
\end{equation}
where $\bm{\Psi}=(\psi_1\cdots, \psi_i,\cdots\psi_N)^{T}$. The nonlinear parameters $g_{ij}$s are usually equal for exact soliton solutions. We would like to derive exact soliton solutions for arbitrary nonlinear parameters by Largangian variational method. We introduce the Lagrangian density as,  $\mathcal{L}=\sum_{i=1}^N\big[\frac{\textrm{i}}{2}(\tilde{\psi_i}^*\partial_t\tilde{\psi_i}-\tilde{\psi_i}\partial_t\tilde{\psi_i}^*)
(1-\frac{a_i^2}{|\tilde{\psi_i}|^2}\delta_{D,i})-\frac{1}{2}|\partial_x\tilde{\psi_i}|^2 -\sum_{j=1}^N\frac{g_{ij}}{2}(|\tilde{\psi_j}|^2-a_j^2 \delta_{D,j})(|\tilde{\psi_i}|^2-a_i^2 \delta_{D,i})\big]$, where
\begin{equation*}
	\delta_{D,i}=\left\{
	\begin{array}{cl}
		0, & \tilde{\psi}_{i}\ \textrm{denotes\ bright soliton}, \\\\
		1, & \tilde{\psi}_{i}\ \textrm{denotes\ dark soliton}, \\
	\end{array}
	\right.
\end{equation*}
to obtain Euler-Lagrangian equations.
In particular, the terms $1-\frac{a_i^2}{|\tilde{\psi_i}|^2}\delta_{D,i}$ and $|\tilde{\psi_i}|^2-a_i^2 \delta_{D,i}$ were introduced in the Lagrangian density for dark soliton \cite{23VD}. Notably, the modified $\mathcal{L}$ corresponds to dynamical equations are,
\begin{equation}\label{xde2}
	\begin{split}
		{\rm{i}}\partial_t \bm{\tilde{\Psi}}=\big[-\frac{1}{2}\partial_{xx}+\sum_{j=1}^{N} g_{ij}(|\tilde{\psi}_j|^2-a_j^2 \delta_{D,j})\big]\bm{\tilde{\Psi}},
	\end{split}
\end{equation}
where $\bm{\tilde{\Psi}}=(\tilde{\psi}_1\cdots, \tilde{\psi}_i,\cdots\tilde{\psi}_N)^{T}$. This is different from Eq. \eqref{xde1}, but the solutions of the two dynamical equations can be transformed to each other by the phase factor $\psi_i/\tilde{\psi}_i=\textrm{e}^{\textrm{i}\theta_i(t)}$, where $\theta_i(t)=-\sum_{j=1}^Ng_{ij}a_j^2t \delta_{D,j}$. After testing many different trial functions with Lagrangian method and the direct numerical simulations, we find that the trial wave function of $i$th component can be written as,
\begin{eqnarray}\label{shitan1}
	\begin{split}
		\tilde{\psi}_{iD}=&\textrm{i}\sqrt{a_i^2-f_i^2(t)}+f_i(t)\tanh\{w_i(t)[x-b(t)]\},\\\\
		\tilde{\psi}_{iB}=&f_i(t)\sech\{w_i(t)[x-b(t)]\}\textrm{e}^{\textrm{i}\{\xi_i(t)+[x-b(t)]\phi_i(t)\}},
	\end{split}
\end{eqnarray}
where $\tilde{\psi}_{iD}$ ($\tilde{\psi}_{iB}$) denotes the wave function of dark (bright) soliton. $f_i$ and $w_i$ describe the amplitude and width of dark (bright) soliton, respectively. $a_i$ is background of dark-component, and the central position of soliton is $b(t)$. $\xi_i$ is time-dependent phase of bright component, and $\phi_i$ is related with the velocity of bright soliton.  Specially, the soliton width parameter $w_i$ are set to be different for different components, although soliton width usually be identical for vector solitons.  These operations are in sharp contrast to the previously used ansatz form, which brings us more constrain conditions on soliton parameters, and finally enable us to obtain exact analytical soliton solutions.  This is our main point for developing variational method.

We obtain the Largrangian ($L$) by substituting Eq. \eqref{shitan1} into $\mathcal{L}$ and integrating over space from $-\infty$ to $+\infty$, which is
\begin{equation}\label{int}
	\begin{split}			L=&\sum_i^N\bigg\{\Big[2f_i^2(t)\phi_i(t)\frac{b'(t)}{w_i(t)}-2f_i^2(t)\frac{\xi_i'(t)}{w_i(t)}-\frac{1}{3}f_i^2(t)w_i(t)-f_i^2(t)\frac{\phi_i^2(t)}{w_i(t)}-\frac{2}{3}g_{ii}\frac{f_i^4(t)}{w_i(t)}\Big]
		\tilde{\delta}_{D,i}+\Big\{-2f_i(t)\sqrt{a_i^2-f_i^2(t)}b'(t)\\&+2 a_i^2\arcsin\Big[\frac{f_i(t)}{a_i}\Big]b'(t)-\frac{2}{3}f_i^2(t)w_i(t)-\frac{2}{3}g_{ii}\frac{f_i^4(t)}{w_i(t)}\Big\}\delta_{D,i}-\sum_{j(j\neq i)}^Ng_{ij}f_i^2(t)f_j^2(t)G_{ij}\bigg\},~~~~~~ (\tilde{\delta}_{D,i}=|\delta_{D,i}-1|).
	\end{split}
\end{equation}
where $G_{ij}=\int^{+\infty}_{-\infty}\big(\pm \sech^2\{w_i(t)[x-b(t)]\}\big)\big(\pm \sech^2\{w_j(t)[x-b(t)]\}\big)dx$, the sign $-~(+)$ corresponding to dark (bright) component. We obtain the equations of motion from Euler-Lagrangian formular $\frac{d}{dt}\big[\frac{\partial L}{\partial \alpha'(t)}\big]=\frac{\partial L}{\partial \alpha(t)}$, where $\alpha(t)$ is variational parameter, $\alpha(t)=f_i(t),w_i(t),w_j(t),\phi_i(t),\xi_i(t),b(t)$. The nontrival equations can be written as following,
\begin{equation}
	\begin{split}
	&\big[g_{11}f_1^2+\sum_{j=2}^N(\frac{3}{2}g_{1j}f_j^2w_1G_{1j}+\frac{3}{2}g_{1j}f_j^2w_1^2\frac{\partial G_{1j}}{\partial w_1})+w_1^2\big]\tilde{\delta}_{D,1}+\big(-g_{11}f_1^2+\sum_{j=2}^N\frac{3}{2}g_{1j}f_j^2w_1^2\frac{\partial G_{1j}}{\partial w_1}+w_1^2\big)\delta_{D,1}=0,\\
	&~~\vdots~~~~~~~~~~~~~~~~~~~~~~~~~~~~~~~~~~~~~~~~~~~~~~~~~~~~~~~~~~~~~~~\vdots~~~~~~~~~~~~~~~~~~~~~~~~~~~~~~~~~~~~~~~~~~~~~~~~~~~~~~~~~~~~\vdots\\
	&\big[g_{ii}f_i^2+\sum_{j\neq i}^{N}(\frac{3}{2}g_{ij}f_j^2w_1G_{ij}+\frac{3}{2}g_{ij}f_j^2w_i^2\frac{\partial G_{ij}}{\partial w_i^2})+w_i^2\big]\tilde{\delta}_{D,i}+\big(-g_{ii}f_i^2+\sum_{j\neq i}^{N}\frac{3}{2}g_{ij}f_j^2w_i^2\frac{\partial G_{ij}}{\partial w_i}+w_i^2\big)\delta_{D,i}=0,\\
	&~~\vdots~~~~~~~~~~~~~~~~~~~~~~~~~~~~~~~~~~~~~~~~~~~~~~~~~~~~~~~~~~~~~~~\vdots~~~~~~~~~~~~~~~~~~~~~~~~~~~~~~~~~~~~~~~~~~~~~~~~~~~~~~~~~~~~\vdots\\
	&\big[\sum_{j=1}^{N-1}(\frac{3}{2}g_{Nj}f_j^2w_NG_{Nj}+\frac{3}{2}g_{Nj}f_j^2w_N^2\frac{\partial G_{Nj}}{\partial w_N})+g_{NN}f_N^2+w_N^2\big]\tilde{\delta}_{D,N}+\big(\sum_{j=1}^{N-1}\frac{3}{2}g_{Nj}f_j^2w_N^2\frac{\partial G_{Nj}}{\partial w_N}-g_{NN}f_N^2+w_N^2\big)\delta_{D,N}=0,
	\end{split}
\end{equation}
where $\frac{\partial G_{ij}}{\partial{w_i}}=-2\int^{+\infty}_{-\infty}\big\{\pm[x-b(t)]\sech^2\{w_i(t)[x-b(t)]\}
\tanh\{w_i(t)[x-b(t)]\}\big\}\big(\pm\sech^2\{w_j(t)[x-b(t)]\}\big)dx$, the sign $-~(+)$ corresponds to dark (bright) component. Then we can get the result of the integral with setting $w_i=w_j=w$.  We consider the $i$th ($j$th) component is dark (bright) soliton, which gives
\begin{equation*}
	\begin{split}
		G_{ij}=&\int^{+\infty}_{-\infty}\big(-\sech^2\{w_i(t)[x-b(t)]\}\big) \big(\sech^2\{w_j(t)[x-b(t)]\}\big)dx\big |_{w_i=w_j}=-\frac{4}{3w},
		\\
		\frac{\partial G_{ij}}{\partial{w_i}}=&-2\int^{+\infty}_{-\infty}\big\{-[x-b(t)]\sech^2\{w_i(t)[x-b(t)]\}
		\tanh\{w_i(t)[x-b(t)]\}\big\}\big(\sech^2\{w_j(t)[x-b(t)]\}\big)dx\big|_{w_i=w_j}=\frac{2}{3w^2},
		\\
		\frac{\partial G_{ij}}{\partial{w_j}}=&-2\int^{+\infty}_{-\infty}\big(\sech^2\{w_i(t)[x-b(t)]\}\big)\big\{-[x-b(t)]\sech^2\{w_j(t)[x-b(t)]\}\tanh\{w_j(t)[x-b(t)]\}\big\}dx\big|_{w_i=w_j}=\frac{2}{3w^2}.
	\end{split}
\end{equation*}
The other cases for $G_{ij}$ and $\frac{\partial G_{ij}}{\partial{w_i}}$ can be calculated in similar ways.  The operations on soliton width parameters are highly non-trivial, which enable us to obtain the exact soliton solutions. Finally, we get the constraint equations of $f_i,w$, as
\begin{eqnarray}
	\left(
	\begin{array}{ccccc}
		g_{11}&\cdots g_{1j}\cdots&g_{1N}&1 \\
		\vdots&\vdots&\vdots&\vdots&\\
		g_{i1}&\cdots g_{ij}\cdots&g_{iN}&1 \\
		\vdots&\vdots&\vdots&\vdots&\\
		g_{N1}&\cdots g_{Nj}\cdots&g_{NN}&1 \\
	\end{array}
	\right)\left(
	\begin{array}{cc}
		\pm f_1^2\\ \vdots\\ \pm f_i^2 \\ \vdots\\ \pm f_N^2\\w^2
	\end{array}
	\right)=0,
\end{eqnarray}
The other parameters are $b'(t)=v,~b(t)=vt,~\phi_i=v$, $\xi_i(t)=\frac{1}{2}(w^2+v^2)t+\theta_i(t)$. The velocity $v=w/f_i\sqrt{a_i^2-f_i^2}$ for dark soliton, and the background constraint between dark soliton is $f_i/f_j$=$a_i/a_j$. The expression of the $i$th component solution is,
\begin{equation}
	\begin{split}
		\psi_{iD}&=\big\{\textrm{i}\sqrt{a_i^2-f_i^2}+f_i\tanh[w(x-vt)]\big\}\textrm{e}^{{\rm{i}}\theta_i(t)},\\
		\psi_{iB}&=f_i\sech[w(x-vt)]\textrm{e}^{\textrm{i}[\frac{1}{2}(w^2+v^2)t+(x-vt)v+\theta_i(t)]}.
	\end{split}
\end{equation}
where $~\theta_i(t)=-\sum_{j=1}^Ng_{ij}a_j^2t\delta_{D,j}$. Noting that all width parameters become identical one $w$ in the expressions for soliton solutions. The exact soliton solutions for $N$-component coupled nonlinear Schr\"{o}dinger equations with arbitrary nonlinear parameters can be given by solving the constraint conditions. Different types vector solitons usually exist in different regions in nonlinear parameters space. Their existence regions can be also clarified by the constrain conditions on soliton parameters.

\section{The explicit expressions of vector solitons in two-component system}\label{2}

The vector solitons for two-component systems can be still classified four families, similar to the integrable  model.
We would like to present the explicit expressions for them as follows.

(i) The exact solution of bright-bright soliton is,
\begin{equation}
	\begin{split}
		\psi_{1B}&=f_1\sech[w(x-vt)]\textrm{e}^{{\textrm{i}}[\frac{1}{2}(w^2+v^2)t+(x-vt)v]},\\
		\psi_{2B}&=f_2\sech[w(x-vt)]\textrm{e}^{{\textrm{i}}[\frac{1}{2}(w^2+v^2)t+(x-vt)v]},
	\end{split}
\end{equation}
where $f_2=\sqrt{\frac{g_{11}-g_{12}}{g_{22}-g_{12}}}f_1,~w=\sqrt{\frac{g_{12}^2-g_{11}g_{22}}{g_{22}-g_{12}}}f_1$.

(ii) The exact solution for dark-bright (DB) or bright-dark (BD) soliton is
\begin{equation}
	\begin{split}
		\psi_{1D}&=\big\{\textrm{i}\sqrt{a_1^2-f_1^2}+f_1\tanh[w(x-vt)]\big\}\textrm{e}^{-{\rm{i}}g_{11}a_1^2t},\\
		\psi_{2B}&=f_2\sech[w(x-vt)]\textrm{e}^{\textrm{i}[\frac{1}{2}(w^2+v^2)t+(x-vt)v]-\textrm{i}g_{21}a_1^2t},
	\end{split}
\end{equation}
where $f_2=\sqrt{\frac{g_{11}-g_{12}}{g_{12}-g_{22}}}f_1,~w=\sqrt{\frac{g_{12}^2-g_{11}g_{22}}{g_{12}-g_{22}}}f_1,~v=\sqrt{\frac{g_{12}^2-g_{11}g_{22}}{g_{12}-g_{22}}}\sqrt{a_1^2-f_1^2}$.
The difference between DB and BD is defined by  total  density  of the two components, which  admits a dip (for DB) or hump (for BD).  Especially, the total density can be uniform when $2g_{12}=g_{11}+g_{22}$, the vector soliton in this case will become the spin soliton.

(iii) The exact solution of dark-dark soliton is,
\begin{equation}
	\begin{split}
		\psi_{1D}&=\big\{\textrm{i}\sqrt{a_1^2-f_1^2}+f_1\tanh[w(x-vt)]\big\}\textrm{e}^{-{\rm{i}}g_{11}a_1^2t-{\rm{i}}g_{12}a_2^2t},\\
		\psi_{2D}&=\big\{\textrm{i}\sqrt{a_2^2-f_2^2}+f_2\tanh[w(x-vt)]\big\}\textrm{e}^{-{\rm{i}}g_{21}a_1^2t-{\rm{i}}g_{22}a_2^2t},
	\end{split}
\end{equation}
where $f_2=\sqrt{\frac{g_{11}-g_{12}}{g_{22}-g_{12}}}f_1,~w=\sqrt{\frac{g_{12}^2-g_{11}g_{22}}{g_{12}-g_{22}}}f_1,~v=\sqrt{\frac{g_{12}^2-g_{11}g_{22}}{g_{12}-g_{22}}}\sqrt{a_1^2-f_1^2},~\frac{a_2}{a_1}=\frac{f_2}{f_1}$.

\section{The explicit expressions of vector solitons in three-component system}\label{3}

We would like to present the explicit expressions for vector soliton solutions in three-component coupled systems as follows.

(i) The exact solution of bright-bright-bright soliton is,
\begin{equation}
	\begin{split}
		\psi_{1B}&=f_1\sech[w(x-vt)]\textrm{e}^{{\textrm{i}}[\frac{1}{2}(w^2+v^2)t+(x-vt)v]},\\
		\psi_{2B}&=f_2\sech[w(x-vt)]\textrm{e}^{{\textrm{i}}[\frac{1}{2}(w^2+v^2)t+(x-vt)v]},\\
		\psi_{3B}&=f_3\sech[w(x-vt)]\textrm{e}^{{\textrm{i}}[\frac{1}{2}(w^2+v^2)t+(x-vt)v]},
	\end{split}
\end{equation}
where $f_2=\sqrt{A_1}f_1,~f_3=\sqrt{A_2}f_1,~w=\sqrt{A_3}f_1$. The parameters $A_1~,A_2,~A_3$ are as follows,
\begin{equation*}
	\begin{split}
		A_1&=\frac{g_{13}^2-g_{11}g_{33}+g_{12}(g_{33}-g_{13})+g_{23}(g_{11}-g_{13})}{g_{23}^2-g_{22}g_{33}+g_{12}(g_{33}-g_{23})+g_{13}(g_{22}-g_{23})},~
		A_2=\frac{g_{12}^2-g_{11}g_{22}+g_{13}(g_{22}-g_{12})+g_{23}(g_{11}-g_{12})}{g_{23}^2-g_{22}g_{33}+g_{12}(g_{33}-g_{23})+g_{13}(g_{22}-g_{23})},\\
		A_3&=\frac{2g_{12}g_{13}g_{23}-g_{13}^2g_{22}-g_{12}^2g_{33}-g_{11}(g_{23}^2-g_{22}g_{33})}{g_{23}^2-g_{22}g_{33}+g_{12}(g_{33}-g_{23})+g_{13}(g_{22}-g_{23})}.
	\end{split}
\end{equation*}
They hold for all vector soltions in three-component models.

(ii) The exact solution of dark-bright-bright soliton is,
\begin{equation}
	\begin{split}
		\psi_{1D}&=\big\{\textrm{i}\sqrt{a_1^2-f_1^2}+f_1\tanh[w(x-vt)]\big\}\textrm{e}^{-{\rm{i}}g_{11}a_1^2t},\\
		\psi_{2B}&=f_2\sech[w(x-vt)]\textrm{e}^{{\textrm{i}}[\frac{1}{2}(w^2+v^2)t+(x-vt)v]-{\textrm{i}}g_{21}a_1^2t},\\
		\psi_{3B}&=f_3\sech[w(x-vt)]\textrm{e}^{{\textrm{i}}[\frac{1}{2}(w^2+v^2)t+(x-vt)v]-{\textrm{i}}g_{31}a_1^2t},
	\end{split}
\end{equation}
where $f_2=\sqrt{-A_1}f_1,~f_3=\sqrt{-A_2}f_1,~w=\sqrt{-A_3}f_1$, and $v=\frac{w}{f_1}\sqrt{a_1^2-f_1^2}$.

(iii) The exact solution of dark-dark-bright soliton is,
\begin{equation}
	\begin{split}
		\psi_{1D}&=\big\{\textrm{i}\sqrt{a_1^2-f_1^2}+f_1\tanh[w(x-vt)]\big\}\textrm{e}^{-{\rm{i}}g_{11}a_1^2t-{\rm{i}}g_{12}a_2^2t},\\
		\psi_{2D}&=\big\{\textrm{i}\sqrt{a_2^2-f_2^2}+f_2\tanh[w(x-vt)]\big\}\textrm{e}^{-{\rm{i}}g_{21}a_1^2t-{\rm{i}}g_{22}a_2^2t},\\
		\psi_{3B}&=f_3\sech[w(x-vt)]\textrm{e}^{{\textrm{i}}[\frac{1}{2}(w^2+v^2)t+(x-vt)v]-{\textrm{i}}g_{31}a_1^2t-{\textrm{i}}g_{32}a_2^2t},
	\end{split}
\end{equation}
where $f_2=\sqrt{A_1}f_1,~f_3=\sqrt{-A_2}f_1,~w=\sqrt{-A_3}f_1,~v=\frac{w}{f_1}\sqrt{a_1^2-f_1^2}$, and $\frac{a_2}{a_1}=\frac{f_2}{f_1}$.

(iv) The exact solution of dark-dark-dark soliton is,
\begin{equation}
	\begin{split}
		\psi_{1D}&=\big\{\textrm{i}\sqrt{a_1^2-f_1^2}+f_1\tanh[w(x-vt)]\big\}\textrm{e}^{-{\rm{i}}g_{11}a_1^2t-{\rm{i}}g_{12}a_2^2t-{\rm{i}}g_{13}a_3^2t},\\
		\psi_{2D}&=\big\{\textrm{i}\sqrt{a_2^2-f_2^2}+f_2\tanh[w(x-vt)]\big\}\textrm{e}^{-{\rm{i}}g_{21}a_1^2t-{\rm{i}}g_{22}a_2^2t-{\rm{i}}g_{23}a_3^2t},\\
		\psi_{3D}&=\big\{\textrm{i}\sqrt{a_3^2-f_3^2}+f_3\tanh[w(x-vt)]\big\}\textrm{e}^{-{\rm{i}}g_{31}a_1^2t-{\rm{i}}g_{32}a_2^2t-{\rm{i}}g_{33}a_3^2t},
	\end{split}
\end{equation}
where $f_2=\sqrt{A_1}f_1,~f_3=\sqrt{A_2}f_1,~w=\sqrt{-A_3}f_1,~v=\frac{w}{f_1}\sqrt{a_1^2-f_1^2}$, and $a_1:a_2:a_3=f_1:f_2:f_3$.